\documentclass{mem}
\usepackage{natbib}\usepackage{txfonts}\usepackage{balance}
\usepackage{graphicx}
\usepackage[a4paper]{hyperref}
\idline{79}{282}
\begin{document}
\def\teff{$T\rm_{eff }$}
\def\kms{$\mathrm {km s}^{-1}$}

\title{
Optical spectroscopy of 3CR sample of radio sources at $z<0.3$}
   \subtitle{}

\author{
S. \,Buttiglione\inst{1} 
\and A. \, Capetti\inst{2}
\and A. \, Celotti\inst{1}
}

\offprints{S. Buttiglione}

\institute{
SISSA/International School for Advanced Studies, Via Beirut 4,
I-34014 Trieste, Italy
\and
INAF - Osservatorio Astronomico di Torino,
Strada Osservatorio 20, 
I-10025 Pino Torinese, Italy
\email{buttigli@sissa.it}
}

\authorrunning{S. Buttiglione et al.}

\titlerunning{3CR optical spectroscopy}

\abstract{We are carrying out a program of optical spectroscopy of the
  complete subsample of the 3CR catalog of radio sources at redshift $z <$
  0.3. The sample consists of 113 3CR sources, comprising FR~I, FR~II radio
  galaxies and Quasars. Complete datasets in other bands are already or will
  be soon available for the whole sample but the optical spectra are sparse
  and inhomogeneous in quality. The observations are carried out at the 3.58m
  Telescopio Nazionale Galileo (TNG, La Palma). More than 100 sources have
  been already observed. We present here the preliminary results on the
  analysis of the high and low resolution spectra. We found that sources can
  be spectroscopically characterized as: High Excitation Galaxies (HEG), Low
  Excitation Galaxies (LEG) and ``Relic'' AGNs.
  This classification is supported by the optical - radio correlations in
  which objects spectroscopically different follow different
  correlations. We conclude that AGNs with the same radio
  power can be fueled with different accretion properties. ``Relic''
  radio-galaxies are characterized by extreme low excitation spectra that we
  interpret as nuclei whose activity has recently turned-off. The
  full spectral catalog will be made available to the scientific community.

}
\maketitle{}

\section{Introduction}

Radio Loud AGNs represent about the 10 per cent of active galaxies,
nevertheless they are among the most powerful objects in the Universe. Since
they are powerful AGNs, they represent an important laboratory to study the
formation and evolution of the nuclear activity.

Our aim is the investigation of their nuclear activity via optical and radio
observations.

\begin{figure*}[t!]
  \resizebox{\hsize}{!}{\includegraphics[clip=true]{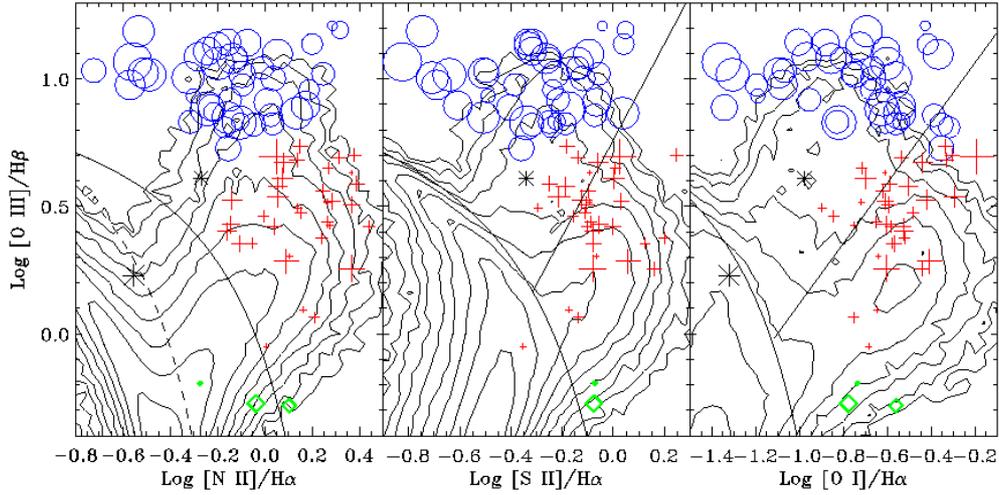}}
  \caption{\footnotesize The diagnostic diagrams are made of pair of emission
    lines ratios: the solid curve represents the separation between AGNs
    (above the line) and star-burst galaxies (below the line). The contour
    lines indicate the distribution of the SDSS galaxies from \citet{k06}.
    Blue circles are High Excitation Galaxies (HEG); red crosses are Low
    Excitation Galaxies (LEG); green diamonds are Relic AGNs; black asterisks
    are unclassified AGNs. The symbol size is proportional to the $L$([O~III])
    luminosity.}
  \label{diag}
\end{figure*}

As emission lines presumably originate from gas photoionized by the
radiation generated via the accretion process, their absolute
luminosities and intensity ratios can provide information on the
accretion itself and thus the physical conditions of the environment in which
the jets are formed.

We focus on the homogeneous and vastly studied 3CR catalog of radio
sources, being unbiased with respect to the nuclear/accretion
properties.

\section{The sample and data analysis}

We consider all sources belonging to the Third Cambridge Revised
catalog (3CR) -- namely all radio sources north of -05 degrees with flux
density higher than 9 Jy at 178 MHz \citep{b62} -- with redshift less
than $z<0.3$, resulting in a complete subsample of 113 objects.

\begin{figure*}[]
%\resizebox{\hsize}{!}{\includegraphics[clip=true,angle=90]{lo3_rc.ps}}
\resizebox{\hsize}{!}{\includegraphics[clip=true,angle=90]{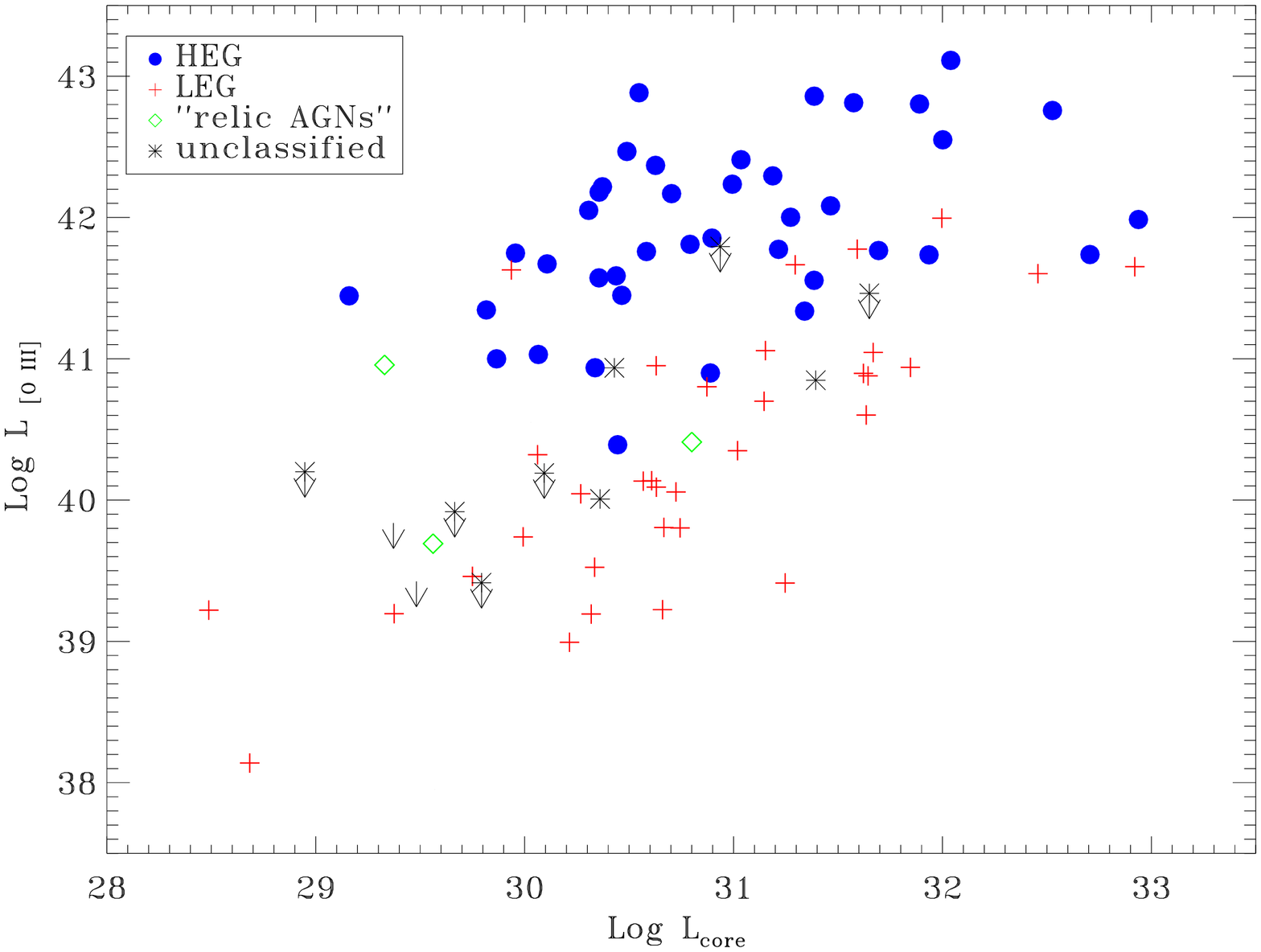}}
\caption{ \footnotesize [O~III]$\lambda$5007\AA\ emission line (erg/s) vs
  radio core power at 5 GHz (erg/s/Hz). Blue circles are High Excitation
  Galaxies (HEG); red crosses are Low Excitation Galaxies (LEG); green
  diamonds are ``relic'' AGNs; black asterisks are spectrally unclassified
  AGNs. Note the vertical offset between HEG and LEG.}
\label{orc}
\end{figure*}

\begin{figure*}[]
\resizebox{\hsize}{!}{\includegraphics[clip=true,angle=90]{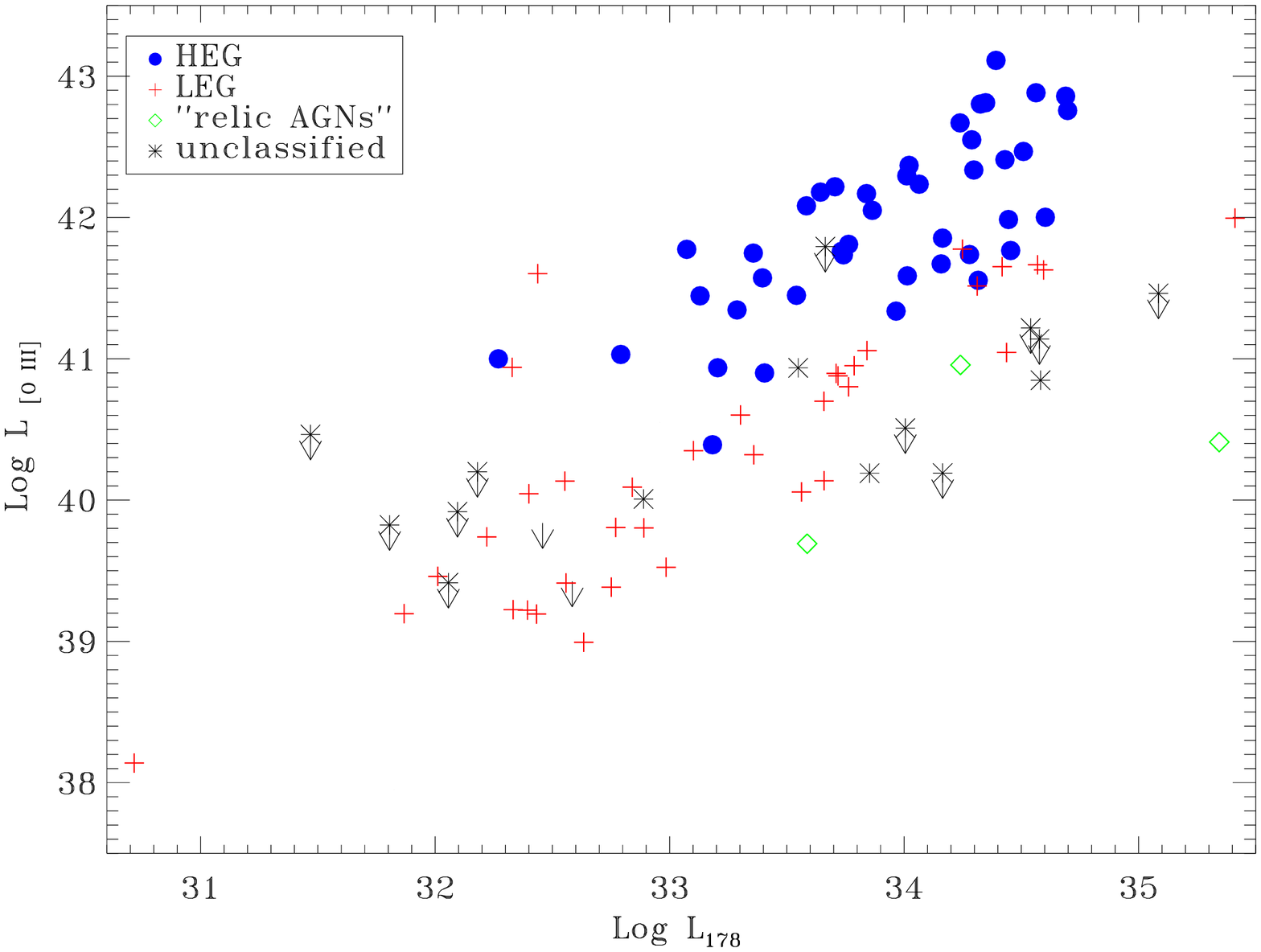}}
\caption{ \footnotesize [O~III]$\lambda$5007\AA\ emission line (erg/s) versus
  extended power at 178 MHz (erg/s/Hz). Blue circles are High Excitation
  Galaxies (HEG); red crosses are Low Excitation Galaxies (LEG); green
  diamonds are Relic AGNs; black asterisks are spectrally unclassified AGNs.
  Note the relatively low [O~III] fluxes of the ``relic'' RG.}
\label{ore}
\end{figure*}

The observations are carried on at the 3.58 m Galileo Telescope (TNG) in
Canary islands.  We use the DOLORES spectrograph with a slit of 2 arc-seconds.
For each target one low resolution spectrum (with the LR-B
grism, 3000 - 8000 \AA, resolution $\sim$20 \AA) and two high resolution
spectra (with the VHR-R grism, 6200 - 7800 \AA, or the VHR-I, 7300 - 8900 \AA,
resolution $\sim$4 \AA) were otained.  Exposure times range between 500 and
1000 sec depending on the redshift of the source.  The IRAF software was used
in order to subtract the bias, divide for the flat field, wavelength and flux
calibrate the rough images.  After correction for the Galaxy extinction, we
considered two wavelengths ranges (of about 1000 \AA) around H$\alpha$ and
H$\beta$ and we subtracted the host galaxy stellar emission using the best fit
single stellar model from the \cite{bc03} library.  Finally the line
intensities were obtained using the {\it specfit} package.

\section{Results}

\subsection{Diagnostic diagrams}
From the narrow emission line fluxes we can create the diagnostic diagrams
(Fig. \ref{diag}): these plots are composed by pairs of emission lines ratios
which are sensitive to the ionizing radiation properties: the
[O~III]$\lambda$5007\AA$/$H$\beta$ is the most important estimator for the
separation of galaxies into star-forming emission and AGN emission; the [N
II]$\lambda$6583\AA$/$H$\alpha$, [S
II]$\lambda\lambda$6717,6731\AA$/$H$\alpha$ and
[OI]$\lambda$6300\AA$/$H$\alpha$ are useful tools for distinguishing low from
high excitation radiation.

In such diagrams it is possible to characterize the sources in at least 3
groups: High Excitation Galaxies (HEG), with typical
[O~III]$\lambda$5007\AA$/$H$\beta$ ratio $>$5; Low Excitation Galaxies (LEG),
with [O~III]$\lambda$5007\AA$/$H$\beta$ $<$5; and a handful of sources with
very low [O~III]$\lambda$5007\AA$/$H$\beta$ ratio ($\sim$0.5), which we
tentatively identify with `relic'' AGNs, as explained below.

\subsection{Optical - radio comparison}
In order to assess any relation between the radio/jet properties and
the nuclear ones, the lines luminosities -- in particular the
[O~III]$\lambda$5007\AA\ -- have been compared with the
radio core emission at 5 GHz $L_{\rm core}$. 

As shown in Fig. \ref{orc}, while
for a fixed $L_{\rm core}$ HEG (blue circles) and LEG (red crosses) are
separated by about a factor of $\sim$ 30 in $L$([O III]), this does
not appear to be connected with radio power. Trends of correlations
between line and radio luminosities appear for both populations, but
there is no transition from LEG to HEG with increasing radio emission.

This suggests that HEG and LEG might be associated to different accretion
mechanisms producing the same AGN manifestation in terms of radio
emission and jet properties.

The same behaviour appear to be present in the relation between [O
III]$\lambda$5007\AA\ and extended radio emission at 178 MHz, as
reported in Fig. \ref{ore}. Again two separate correlations can be
distinguished: HEG (blue cirlces) and LEG (red crosses) differ by about a
factor of $\sim$ 10 in $L$([O III]) over the entire radio power range.

\subsection{A new spectroscopic class associated to relic AGNs?}

Interestingly, the (few) sources spectroscopic defined above as
``relic'' galaxies reveal to have among the lowest levels of $L$([O
III]) emission. They are thus characterized by small [O
III]$\lambda$5007\AA$/$H$\beta$ ratio (Fig. \ref{diag}), weak [O
III]$\lambda$5007\AA\ luminosity and small ratio of core vs exteded
radio emission (see Figs. \ref{orc} and \ref{ore}).

From these findings our best hypothesis for these sources is that they
are galaxies in which the nuclear activity has ``recently'' turned
off: this could explain their weak/absent radio core but still the
presence of the extended radio emission which would react to changes
of central activity on much longer timescales.  The weak line emission
and extreme line ratios can just reflect the response of the Narrow
Line Region gas -- on $<$ kpc scales -- to a vanishing photoionizing
nuclear radiation on an intermediate timescale between core
and large scale radio emission.
The actual number density of such ``relic'' sources would provide a
crucial piece of information on the duty-cycle of activity of the
engine and in turn on the co-evolution and relationship between the
host galaxy and the active nucleus.

\bibliographystyle{aa}

\end{document}